\documentstyle[11pt]{article}

\begin{document}

\title{SIMILARITY AND CONSIMILARITY OF ELEMENTS IN REAL CAYLEY-DICKSON
ALGEBRAS}

\bigskip
\author{Yongge   Tian \\
Department of Mathematics and Statistics \\
Queen's University \\
Kingston,  Ontario,  Canada K7L 3N6\\
{\tt e-mail:ytian@mast.queensu.ca}}
\maketitle

\bigskip

\noindent {\bf Abstract.}  \ {\small Similarity and consimilarity of elements in the real quaternion, octonion, and sedenion algebras, as well as in the general real Cayley-Dickson algebras are considered by solving the two fundamental
 equations $ ax = xb $ and $ax = \overline{x}b$ in these algebras. Some
 consequences are also presented. \\


} \\

\bigskip

\noindent {\bf 1. \ Introduction } \\

\noindent We consider in the article how to establish the concepts of similarity and
consimilarity for elements in the real quaternion, octonion and sedenion algebras,
 as well as in the $2^n$-dimensional real Cayley-Dickson algebras. This consideration 
is motivated by some recent work on eigenvalues and eigenvectors, as well as 
similarity of matrices over the real quaternion and octonion algebras(see \cite{DM1} and \cite{Zh}). In order to establish a set of complete theory  on 
eigenvalues and eigenvectors, as well as similarity of matrices over 
the quaternion, octonion, sedenion algebras, as well as the general real Cayley-Dickson algebras, one must first consider a basic problem---
how to characterize similarity of elements in these algebras, which leads us to the work in this article.

\medskip

Throughout ${\cal H}, \ {\cal O},$ and ${\cal S }$ denote the real
quaternion, octonion, and sedenion algebras, respectively; ${\cal A}_{n}$ 
denotes the $2^n$-dimensional real Cayley-Dickson algebra,
 ${\cal R}$ and ${\cal C}$ denote the real and complex number fields, respectively.  It is well known that ${\cal A}_0 = {\cal R}$, ${\cal A}_1 = {\cal C}$,  ${\cal A}_2 = {\cal H}$,
 ${\cal A}_3 = {\cal O}$, and  ${\cal A}_4 = {\cal S}$. 

\medskip

As is well-known, the real quaternion algebra  ${\cal H}$ is a four-dimensional associative 
division algebra over the real number field ${\cal R}$ with its basis
$ 1, \ i, \ j, \ k $ satisfying the multiplication rules
$ i^2 = j^2 = k^2 = -1,$ $ ij = -ji=k,$ $ jk = -kj = i,$ $ ki = -ik =j.$  
The elements in ${\cal H}$ take the form
 $ a = a_0 + a_1i + a_2j + a_3k,$ where $ a_0$---$a_3  \in {\cal R}$, which
 can  simply be written as $ a = {\rm Re\,}a +  {\rm Im\,}a,$ where
 ${\rm Re\,}a = a_0$ and ${\rm Im\,}a =  a_1i + a_2j + a_3k.$ The {\em conjugate} of $ a $ is defined to be $  \overline{a} = a_0 - a_1i - a_2j - a_3k
 = {\rm Re\,}a -  {\rm Im\,}a,$ which satisfies
 $ \overline{\overline{a}} = a, \ \overline{a+b} =  \overline{a} +
 \overline{b}$ and $ \overline{ab} =   \overline{b}\overline{a}$ for all
 $ a, \ b  \in {\cal H}.$  The {\em norm} of $ a $ is defined to be
 $ |a | = \sqrt{a\overline{a}} = \sqrt{\overline{a}a} =
 \sqrt{ a_0^2 + a_1^2 + a_2^2 + a_3^2 }.$ Some basic operation properties
 on quaternions are listed below
$$ 
a^2 - 2({\rm Re\,}a)a + |a|^2 = 0,  \qquad({\rm Im\,}a)^2
= -| {\rm Im\,}a|^2,   \eqno(1.1)
$$ 
$$ 
|ab| = |a||b|,  \eqno(1.2)
$$ 
$$
a^{-1} = \frac{ \overline{a}}{ |a|^2},    \eqno(1.3)
$$   
$$ 
{\rm Re\,}(ab) = {\rm Re\,}(ba).  \eqno(1.4) 
$$ 

As for octonions, they can be defined by the Cayley-Dickson process as
an extension of quaternions as follows
$$ 
a = a' + a''e,
$$
where $a', \  a'' \in  {\cal H}$, and the addition and multiplication for
any  $ a = a' + a''e, \ b = b' + b''e \in {\cal O}$ are 
$$ 
a  + b = ( a' + a''e)  + ( b' + b''e ) =  ( a' + b') +   ( a'' +  b'')e,
 \eqno(1.5)
$$  
$$ 
ab = (a' + a''e )(b' + b''e) = ( a'b'  -   \overline{b''}a'' )
 + ( b''a' +  a''\overline{b'})e.   \eqno(1.6)
$$ 
In that case, ${\cal O}$  is spanned as an eight-dimensional non-associative
but alternative division algebra over the real number field
 ${\cal R}$ with its canonical basis as follows
$$
 1, \ \ \  e_1 =i, \ \ \  e_2 = j, \ \ \  e_3 = k, \ \ \  e_4 =e, \ \ \
 e_5 = ie, \ \  e_6 = je, \ \ \  e_7 = ke.    \eqno(1.7)
$$
The multiplication table for the basis can be derived from
(1.6)( see, e.g., \cite{Sc2}), but we omit it here. In term of  (1.7),
all elements of  ${\cal O}$ take the form
$$  
a = a_0 + a_1e_1 + \cdots + a_7e_7,
$$ 
where $ a_0$---$a_7 \in {\cal R}$, which can simply be written as
$a = {\rm Re\,}a +  {\rm Im\,}a,$
where $ {\rm Re\,}a = a_0.$ The {\em conjugate} of $ a =a' + a''e $ is defined to be
$ \overline{a} = \overline{a'} - a''e = {\rm Re\,}a -
{\rm Im\,}a,$ which satisfies $ \overline{\overline{a}} = a,
\ \overline{a+b} =  \overline{a} +  \overline{b}$ and $ \overline{ab} =
 \overline{b}\overline{a}$  for all $ a, \ b  \in {\cal O }.$   The {\em norm} of
 $ a $ is defined to be $ |a | = \sqrt{a\overline{a}} =
 \sqrt{\overline{a}a} =  \sqrt{ a_0^2 + a_1^2 + \cdots + a_7^2 }. $ Some
 basic operation properties used in the sequel on octonions are listed below
$$ 
a(ab) = a^2b,  \qquad (ba)a = ba^2,   \eqno(1.8)
$$ 
$$ 
(ab)a = a(ba):= aba,  \ \ \ (ab)a^{-1} = a(ba^{-1}) := aba^{-1},
\eqno(1.9)
$$ 
$$
a^{-1} = \frac{ \overline{a}}{ |a|^2},    \eqno(1.10) 
$$
$$ 
a^2 - 2({\rm Re\,}a)a + |a|^2 = 0,  \qquad   ({\rm Im\,}a)^2
= -| {\rm Im\,}a|^2,   \eqno(1.11)
$$ 
$$ 
|ab| = |a||b|,  \eqno(1.12)
$$ 
$$ 
{\rm Re\,}(ab) = {\rm Re\,}(ba).  \eqno(1.13)
$$ 

By the Cayley-Dickson process, the sedenion algebra ${\cal S}$ is an
extension of the octonion algebra ${\cal O}$  with adding a new generator
$ \varepsilon$ in it. The elements in $ {\cal S}$ take the form
$$ 
a = a' + a'' \varepsilon,  
$$
where $a', \  a'' \in  {\cal O}$, and the addition and multiplication for
any  $ a = a' + a''\varepsilon, \
  b = b' + b'' \varepsilon \in {\cal S}$ are defined by
$$ 
a  + b = ( a' + a'' \varepsilon)  + ( b' + b'' \varepsilon )
=  ( a' + b') +   ( a'' +  b'') \varepsilon,
 \eqno(1.14)
$$  
$$ 
ab = (a' + a'' \varepsilon)(b' + b'' \varepsilon)
= ( a'b'  -   \overline{b''}a'' ) + (  b''a' +
a''\overline{b'}) \varepsilon.   \eqno(1.15)
$$ 
In that case, ${\cal S}$  is spanned as a sixteen-dimensional nonassociative
 algebra over ${\cal R}$ with its canonical basis as follows
$$
 1, \ \ \  e_1, \ \ \ \cdots, \ \ \  e_7, \ \ \   e_8 = \varepsilon, \ \ \
  e_9 =e_1\varepsilon, \ \ \   \cdots, \ \ \ e_{15} =e_7\varepsilon,
   \eqno(1.16)
$$
where $1, \ e_1$---$e_7$ are the  canonical basis of ${\cal O}$. The multiplication
table for this basis can be found in \cite{SL}. In term of (1.16), all elements of
  ${\cal S}$ can be written as
$$  
a = a_0 + a_1e_1 + \cdots + a_{15}e_{15},
$$ 
where $ a_0$---$a_{15}  \in {\cal R}$, or simply  $  a = {\rm Re\,}a +
{\rm Im\,}a,$ where $ {\rm Re\,}a = a_0.$ The {\em conjugate} of $ a $ is defined
to be $ \overline{a} = \overline{a'} - a'' \varepsilon
= {\rm Re\,}a -  {\rm Im\,}a,$ which satisfies $  \overline{\overline{a}}
= a$ and $ \overline{a+b} =  \overline{a} +  \overline{b},  \  \overline{ab} 
=   \overline{b}\overline{a} $ for all $ a, \ b  \in {\cal S}.$  The {\em norm}
 of $ a $ is defined to be $  |a | = \sqrt{a\overline{a}} = \sqrt{\overline{a}a} =
 \sqrt{ a_0^2 + a_1^2 + \cdots + a_{15}^2 }.$ It is well known that ${\cal S}$ is a non-commutative, non-alternative, non-composition
 and non-division algebra, but it is a  power-associative, flexible and
 quadratic algebra over $ {\cal R}$, namely,
$$ 
aa^2 = a^2a,  \qquad  a(aa^2) = a^2a^2,     \eqno(1.17)
$$ 
$$ 
(ab)a = a(ba):= aba,   \eqno(1.18)
$$ 
$$ 
a^2 - 2({\rm Re\,}a)a + |a|^2 = 0,  \ \ \  ({\rm Im\,}a)^2 = -| {\rm Im\,}a|^2.   \eqno(1.19)
$$ 
In addition
$$
a^{-1} = \frac{ \overline{a}}{ |a|^2},     \eqno(1.20)
$$
$$ 
{\rm Re\,}(ab) = {\rm Re\,}(ba),  \eqno(1.21)
$$
$$
(ab)\overline{a} = a(b\overline{a}) := ab\overline{a}, \ \ \ (ab)a^{-1} = a(ba^{-1}):= aba^{-1}. \eqno(1.22)
$$ \\

\noindent {\bf 2. Similarity and consimilarity of quaternions}
 \\

\noindent In this section, we fist solve two basic equations $ ax= xb$ and
$ ax = \overline{x}b$, and then introduce the concepts of similarity and
consimilarity for elements in $ {\cal H }$. In addition, we shall also present
some interesting consequences. 

\medskip

\noindent {\bf  Theorem  2.1.} \ {\em  Let $a = a_0 + a_1i + a_2j + a_3k$ and
$b = b_0 + b_1i + b_2j + b_3k $ be two given quaternions. Then

{\rm (a)\cite{Bre} \cite{Zh}} \ The linear equation
$$
 ax = xb   \eqno (2.1) 
$$ 
has a nonzero solution for $ x \in {\cal H}$ if and only if
$$
 a_0 = b_0,  \qquad  and  \qquad  | {\rm Im\,}a | = | {\rm Im\,}b |.
 \eqno (2.2)
$$ 

{\rm (b)} \ In that case$,$ the general solution of Equation {\rm (2.1)} is 
$$ 
 x = ({\rm Im\,}a )p + p({\rm Im\,}b),  \eqno (2.3)
$$ 
or equivalently $ x = ap - p\overline{b}, $ where $ p \in {\cal H }$ is
arbitrary.

{\rm (c)}\ In particular$,$ if $ b \neq \overline{a},$ i.e.$,$ ${\rm Im}\,a
+ {\rm Im}\,b \neq 0,$  then Equation {\rm (2.1)} has a solution as follows
$$
 x = \lambda_1( \: {\rm Im\,}a  + {\rm Im\,}b\:) + 
\lambda_2 [\: |{\rm Im\,}a |\:|{\rm Im\,}b| -( {\rm Im\,}a )({\rm Im\,}b ) \:],  \eqno (2.4)
$$ 
where $ \lambda_1, \ \lambda_2 \in {\cal R }$ are arbitrary.  If
$b = \overline{ a },$ then the general solution to Equation {\rm (2.1)}  can be
written as
$$
 x =  x_1i + x_2j + x_3k, \eqno (2.5)
$$
where  $x_1, \ x_2 $ and $x_3 $ satisfy $ a_1x_1 + a_2x_2 + a_3x_3 = 0.$
}

\medskip

\noindent {\bf Proof.} \ Assume first that (2.1) has a nonzero solution
$ x\in {\cal H }$ . Then by (1.2) and (1.4) we find
$$ 
ax = xb  \Longrightarrow |ax| = |xb| \Longrightarrow  |a||x| = |x||b|
 \Longrightarrow 
 |a| = |b| 
$$
and   
$$ 
ax = xb  \Longrightarrow  a = xbx^{-1} \Longrightarrow
 {\rm Re\,}a = {\rm Re\,}( xbx^{-1}) = {\rm Re\,}(bx^{-1}x) = {\rm Re\,}b, 
$$ 
which are equivalent to (2.2). Conversely, substituting (2.3) into
$ ax - xb $, we obtain
\begin{eqnarray*} 
ax- xb  & = & ( a_0 + {\rm Im\,}a )[ ({\rm Im\,}a)p + p({\rm Im}\,b )] -
[ ({\rm Im\,}a)p + p({\rm Im}\,b )]( b_0 + {\rm Im}\,b) \\
 & = & ( a_0 -b_0 )x + ({\rm Im\,}a )^2p + ( {\rm Im\,}a)p({\rm Im\,}b ) - 
({\rm Im\,}a)p({\rm Im\,}b ) - p({\rm Im\,}b )^2 \\
& = &  ( a_0 -b_0 )x + ( |{\rm Im}\,b|^2 - |{\rm Im}\,a|^2 )p.    
\end{eqnarray*}   
Under (2.2) the right-hand side of the above equality is zero. Thus (2.3)
is a solution of (2.1). Next suppose that $ x_0$ is any solution of (2.1)
under (2.2), and let
$$
 p = - \frac{ ({\rm Im}\,a)x_0 }{ 2 |{\rm Im}\,a|^2}
 =  - \frac{x_0({\rm Im}\,b)}{ 2 |{\rm Im}\,b|^2}.
$$
Then (2.3) becomes 
$$ 
x = - \frac{({\rm Im}\,a)^2x_0}{ 2 |{\rm Im}\,a|^2 } -
\frac{x_0({\rm Im}\,b)^2}{ 2 |{\rm Im}\,b|^2} =
  \frac{1}{2}x_0 + \frac{1}{2}x_0 = x_0,
$$ 
which shows that any solution to (2.1) can be represented by (2.3). Thus
(2.3) is the general solution of (2.1) under (2.2). If setting $ p = 1 $ in
(2.3), then we know that  $ x_1 = {\rm Im\,}a  + {\rm Im\,}b $ is a
special solution to (2.1), and if setting $ p =  - {\rm Im}\,b$ in (2.3),
then we get $ x_2 = |{\rm Im\,}a |\:|{\rm Im\,}b| -
( {\rm Im\,}a )({\rm Im\,}b )$, another special solution to (2.1). Thus
 $ x = \lambda_1 x_1 + \lambda_1 x_2$, where $ \lambda_1, \ \lambda_2 \in
 {\cal R }$ are arbitrary,  is also a solution to
(2.1) under (2.2) and $ b \neq \overline{ a }.$ If $ b = \overline{ a },$
then (2.1) is $ ax = x\overline{a}$, namely
$$ 
2 ({\rm Re\,}x) ( {\rm Im\,}a) +  ({\rm Im\,}a)({\rm Im\,}x) + ({\rm Im\,}x)({\rm Im\,}a) = 0,
$$ 
which is also equivalent to 
$$ 
{\rm Re\,}x = 0,  \ \ and \ \   a_1x_1 + a_2x_2 + a_3x_3 = 0.
$$ 
Thus we have (2.5).  \qquad $ \Box $ 

\medskip

An equivalent statement for (2.1) to have a nonzero solution is that
$ a $ and $ b $ are {\em similar}, which is written by $ a \sim b $. A
direct consequence of Theorem 2.1 is given below. 

\medskip

\noindent {\bf Corollary 2.2.} \ {\em Let $ a \in { \cal H }$  given with $ a \notin
{\cal R}.$ Then the equation
$$
 ax = x( \: {\rm Re}\,a  + |{\rm Im}\,a |i \: ) \eqno (2.6)
$$
always has a nonzero solution$,$  namely$,$  $  a  \ \sim \  {\rm Re}\,a  +
|{\rm Im}\,a |i,$ and the general  solution to Equation {\rm (2.6)} is
 $$
x = ({\rm Im\,}a)p + |{\rm Im\,}a|pi,    \eqno (2.7)
$$
where $ p \in {\cal H }$ is arbitrary. In particular$,$   if $ a \notin
{\cal C},$  then a  solution of Equation {\rm(2.6)} is
$$
  x = \lambda_1[ \ |{\rm Im}\,a |i  + {\rm Im}\,a \ ] + \lambda_2 [ \ |{\rm Im}\,a | -( {\rm Im}\,a )i \ ],
$$
where $ \lambda_1, \  \lambda_2 \in {\cal R }$ are arbitrary. If 
$ a = a_0 + a_1i $ with $ a_1 < 0,$  then the general solution to {\rm (2.6)} is 
$$ 
x = x_2j + x_3 k, \ \ \  for \ all \  x_2, \ x_3 \in {\cal R }.
$$   }

Through (2.6), one can easily find out powers and $n$th roots of quaternions.
 This topic was previously examined in \cite{Bra},  \cite{Ni2} and 
 \cite{Sa}. 

\medskip  

\noindent {\bf  Theorem  2.3.} \ {\em  Let $a = a_0 + a_1i + a_2j + a_3k$ and
$b = b_0 + b_1i + b_2j + b_3k $ be two given quaternions. Then the equation
$$
 ax = \overline{x}b   \eqno (2.8) 
$$ 
has a nonzero solution for $ x \in {\cal H }$ if and only if  
$$
|a| = |b |. \eqno (2.9) 
$$ 
In that case$,$  if $ a + \overline{b} \neq 0,$ then Equation {\rm (2.8)} has a solution as follows
$$
 x = \lambda(  \overline{a} + b ),   \eqno (2.10)
$$ 
where $ \lambda \in {\cal R }$ is arbitrary. If $ a + \overline{b} = 0,$
then the general solution to Equation {\rm (2.8)} is can be written as
$$
 x =  x_0 +  x_1i + x_2j + x_3k, \eqno (2.11)
$$
where  $x_0$---$x_3 $ satisfy $  a_0x_0 - a_1x_1 - a_2x_2 - a_3x_3 = 0.$  }

\medskip

\noindent {\bf Proof.} \  Suppose first that (2.8) has a nonzero solution
$ x \in { \cal H }$. Then by (1.2) we get
$$ 
ax = xb  \Longrightarrow  |ax| = |xb|  \Longrightarrow  |a||x| = |x||b|
 \Longrightarrow 
 |a| = |b|. 
$$
Conversely, we let $ x_1 =\overline{a} + b$. Then
\begin{eqnarray*} 
ax_1- \overline{x}_1b & = & a(\overline{a} + b ) -  ( a +  \overline{b})b  \\
 & = & a\overline{a} + ab - ab +  \overline{b}b = |a|^2 -  |b|^2.
\end{eqnarray*}   
Thus (2.10) is a solution to (2.8) under (2.9). If $ a + \overline{b} = 0$, then (2.8) is equivalent to 
$ax +  \overline{ax} = 0$, i.e., $ {\rm Re\,}(ax)= 0$. Thus we have (2.11).     \qquad $\Box $ 

\medskip

Based on the equation in (2.8) we can also extend the concept of consimilarity on complex matrices ( see,
 e.g., \cite{HJ}) to quaternions. Two quaternions $ a$ and $ b $ are said to
 be {\em consimilar} if there is a nonzero $ p \in {\cal H }$  such that 
$ a = \overline{p}bp^{-1}$. By Theorem 2.3, we immediately know that two
 quaternions are  consimilar  if and only if their norms are identical.
 Thus  the consimilarity defined here  is also an equivalence relation on
 quaternions. 

\medskip

\noindent {\bf Corollary 2.4.} \ {\em  Any quaternion $ a $ with $ a \notin
{\cal R }$ is consimilar to its norm$,$
namely$,$  $ a =  \overline{p}|a|p^{-1},$ where $ p = |a| + \overline{a} $. }

\medskip

\noindent {\bf Corollary 2.5.}  \ {\em Let   $a, \ b  \in {\cal H }$ be given  with $ a, \ b \notin {\cal R}.$  Then $a|a|^{-1} $ and $ b|b|^{-1} $ are consimilar. }

\medskip

An interesting consequence of Corollary 2.4 is given below.

\medskip

\noindent {\bf Corollary 2.6.} \ {\em  Let $ a \in {\cal H }$ be given with $ a  \notin
{\cal R }$.  Then the quadratic equation $ x^2 = a $ has two quaternion
solutions as follows
$$ 
x = \pm \frac{|a|^{\frac{1}{2}}( \, |a| + a \,)}{  | \, |a| + a \,|}
= \pm ( \, \lambda_0 + \lambda_1 a \,), \eqno (2.12)
$$
where $ \lambda_0 =   \frac{ |a|^{\frac{3}{2}} }{| \, |a| + a \,| }$
 and $\lambda_1 = \frac{|a|^{\frac{1}{2}}}{ | \, |a| + a \,|}.$ } 

\medskip

\noindent {\bf Proof.} \ By Corollary 2.4, we can write $ a $ as  $ a = |a|
 \overline{p}p^{-1}$, where $ p = |a| + \overline{a}.$ Thus by (1.3),
 we have
$$ 
a =  |a| \frac{ \overline{p}^2}{|p|^2} = \left( \,|a|^{\frac{1}{2}}
\frac{\overline{p}}{|p|} \,\right)^2,
$$
which shows that the two quaternions in (2.12) are the solutions to $ x^2 = a$.
\qquad $ \Box $ 

\medskip

The above result can also be restated that any quaternion $ a $ with $ a
\notin {\cal R } $ has two square roots as in (2.12). Based on the result
in Corollary 2.6, solutions can also explicitly be derived  for  some other 
simple quadratic equations over ${\cal H}$, such as,  $ xax = b, 
\ x^2 + bx + xb + c = 0 $ and $ x^2 + xb + c = 0 $ with $ bc = cb $. 

\medskip

Based on  the results in Theorem 2.1, we can also solve the quaternion 
equation $ \overline{x}ax = b$, which was exmined previouly in \cite{Po}. 

\medskip

\noindent {\bf Theorem 2.7.} \ {\em  Let $ a, \ b  \in {\cal H }$ be given with $ a, \ b \notin
{\cal R }$.  Then the equation $ \overline{x}ax = b$ is solvable if and only 
if there is a $\lambda \in {\cal R }$ with $\lambda > 0$ such that 
$$ 
{\rm Re\,}a = \lambda {\rm Re\,}b, \ \ \ and  \ \ \ |{\rm Im\,}a| 
= \lambda |{\rm Im \,}b|. \eqno (2.13)
$$
In that case$,$ a solution to $ \overline{x}ax = b$  can be written 
as 
$$ 
x = \frac{ ({\rm Im\,}a)p + p({\rm Im\,}b)}{\sqrt{\lambda}\, |\, ({\rm Im\,}a)p 
+ p({\rm Im\,}b) \,|},  \eqno (2.14) 
$$ 
where $ p \in {\cal H}$ is arbitrarily chosen such that 
$ ({\rm Im\,}a)p + p({\rm Im\,}b) \neq 0$.}  

\medskip

\noindent {\bf Proof.} \ Notice that $x^{-1} =  \overline{x}/|x|^2$ for a nonzero quaternion $x$.  Thus the equation $ \overline{x}ax = b$ can equivalently be written as 
$$ 
ax = \frac{x}{|x^2|}b. \eqno (2.15) 
$$ 
If (2.15) is solvable for $x$, then 
$$
{\rm Re\,}a = \frac{1}{|x|^2}{\rm Re\,}b, \ \ \ {\rm and} \ \ \ 
|{\rm Im\,}a| = \frac{1}{|x|^2}|{\rm Im\,}b|
$$
by Theorem 2.1(a), which implies (2.13). Conversely if (2.13) holds,  it is easy to verify that (2.14) satisfies (2.15). \qquad $\Box$
      
\medskip   

Without much effort, all the above results can be extended to the octonion
algebra, which are given in the next section. \\

\noindent {\bf 3. Similarity and consimilarity of octonions}
 \\

\noindent {\bf  Theorem 3.1.} \ {\em  Let $a = a_0 + a_1e_1 + \cdots +  a_7e_7$ and
$ b = b_0 + b_1e_1 + \cdots +  b_7e_7 $ be two given octonions. Then  

{\rm (a)}  \ The linear equation
$$
 ax = xb   \eqno (3.1) 
$$ 
has a nonzero solution for $ x \in {\cal O }$ if and only if  
$$
 a_0 = b_0, \qquad   and \qquad  | {\rm Im\,}a | = | {\rm Im\,}b |.
 \eqno (3.2)
 $$ 

{\rm (b)} \ In that case$,$ if $ b \neq \overline{a},$ i.e.$,$ $ {\rm Im}\,a
+ {\rm Im}\,b \neq 0,$  then the general solution of  Equation {\rm (3.1)} can be
expressed as
$$ 
 x = ({\rm Im\,}a )p + p({\rm Im\,}b),  \eqno (3.3)
$$ 
where $p$ is arbitrarily chosen in ${\cal A }(a, \, b ),$  the subalgebra generated by
 $ a $ and $ b$. In particular$,$ Equation {\rm(3.1)} has a solution as follows
 $$
 x = \lambda_1( \: {\rm Im\,}a  + {\rm Im\,}b\:) + \lambda_2
 [\: |{\rm Im\,}a |\:|{\rm Im\,}b| -( {\rm Im\,}a )({\rm Im\,}b ) \:],
  \eqno (3.4)
$$ 
where $ \lambda_1, \ \lambda_2 \in {\cal R }$ are arbitrary. 

{\rm (c)} \ If $ b = \overline{a},$ then the general solution of
 Equation {\rm (3.1)} is
$$
 x =  x_1e_1 + x_2e_2  + \cdots +  x_7e_7, \eqno (3.5)
$$
where  $x_1$---$x_7$ satisfy $ a_1x_1 + a_2x_2  + \cdots + a_7x_7 = 0.$
  }  

\medskip  

\noindent {\bf Proof.} \ Assume first that (3.1) has a nonzero solution $ x \in
{\cal O }$. Then by (1.8), (1.9) and (1.12), (1.13) we find
$$ 
ax = xb \ \Longrightarrow \ |ax| = |xb|   \Longrightarrow 
 |a||x| = |x||b|  \Longrightarrow  |a| = |b|
$$
and   
$$ 
ax = xb \ \Longrightarrow  \ a = xbx^{-1} \Longrightarrow
 {\rm Re\,}a = {\rm Re\,}( xbx^{-1}) =
{\rm Re\,}(bx^{-1}x) = {\rm Re\,}b, 
$$ 
which are equivalent to (3.2). Conversely, note that $ p
\in {\cal A }(a, \, b )$ in (3.3). The products of $ a, b $ with $ p $
 are associative. Thus $ x $ in (3.3) and $ a, b $ satisfy
\begin{eqnarray*} 
ax- xb & = & ( a_0 + {\rm Im\,}a )[ ({\rm Im\,}a)p + p({\rm Im}\,b )]
- [ ({\rm Im\,}a)p + p({\rm Im}\,b )]( b_0 + {\rm Im}\,b) \\
 & = & ( a_0 -b_0 )x + ({\rm Im\,}a )^2p + ( {\rm Im\,}a)p({\rm Im\,}b ) - 
({\rm Im\,}a)p({\rm Im\,}b ) - p({\rm Im\,}b )^2 \\
& = &  ( a_0 -b_0 )x + ( |{\rm Im}\,b|^2 - |{\rm Im}\,a|^2 )p ,    
\end{eqnarray*}   
Under (3.2) the right-hand sides of the above equality is zero. Thus (3.3)
  is a  solution of (3.1).
Next suppose that $ x_0$ is any solution of (3.1) under (3.2),
then it must belong to $ {\cal A }(a, \, b )$ because (3.1) is linear.
 Now let
 $$
 p = - \frac{ ({\rm Im}\,a)x_0 }{ 2 |{\rm Im}\,a|^2}
 =  - \frac{x_0({\rm Im}\,b)}{ 2 |{\rm Im}\,b|^2},
$$
 in (3.3). Then $ p \in  {\cal A }(a, \, b )$ and  (3.3) becomes 
$$ 
x = - \frac{({\rm Im}\,a)^2x_0}{ 2 |{\rm Im}\,a|^2} -
\frac{x_0({\rm Im}\,b)^2}{ 2 |{\rm Im}\,b|^2 } =
  \frac{1}{2}x_0 + \frac{1}{2}x_0 = x_0,
$$ 
which shows that any solution to (3.1) can be represented by (3.3).
 Thus (3.3) is the general solution of
(3.1) under (3.2). If setting $ p = 1 $ in (3.3), then we know that
$ x_1 = {\rm Im\,}a  + {\rm Im\,}b $ is a special solution to (3.1),
and if setting $ p = -{\rm Im}\,b$ in (3.3), then we get $ x_2 =
 |{\rm Im\,}a |\:|{\rm Im\,}b| -( {\rm Im\,}a )({\rm Im\,}b )$, another
  special  solution to (3.1). Thus $ x = \lambda_1 x_1 + \lambda_1 x_2$,
   where $ \lambda_1, \ \lambda_2 \in {\cal R }$ are arbitrary,
    is also a solution to
(3.1) under (3.2) and  $ b \neq \overline{ a }.$  If
 $ b = \overline{ a },$ then (3.1) is $ ax = x\overline{a}$, namely
$$ 
2 ({\rm Re\,}x) ({\rm Im\,}a) +  ({\rm Im\,}a)({\rm Im\,}x)
+ ({\rm Im\,}x)({\rm Im\,}a) = 0,
$$ 
which is also equivalent to 
$$ 
{\rm Re\,}x = 0,  \ \ {\rm and} \ \   a_1x_1 + a_2x_2 + \cdots +  a_7x_7 = 0.
$$ 
Thus we have (3.5).  \qquad  $\Box $

\medskip  

We guess that (3.4) is equivalent to (3.3), but fail to give a satisfactory 
proof. 

\medskip

Based on the equation (3.1), we can define similarity of octonions.
 Two octonions $ a $ and $ b$ are said to be {\em similar} if there is a nonzero
 $  p \in {\cal O }$  such that $ a = pbp^{-1}$, which  is
  written as
$ a \sim b$. By Theorem 3.1, we immediately know that two octonions
 are similar  if and only if  $  {\rm Re\,}a  = {\rm Re\,}b$
 and $ | {\rm Im\,}a | = | {\rm Im\,}b |$. Thus  the similarity
 defined here is also an equivalence relation on octonions.

\medskip   

 A direct consequence of Theorem 3.1 is given below. 

\medskip

\noindent {\bf Corollary 3.2.}  \ {\em For any  $ a \in { \cal O }$ with $ a
\notin {\cal C},$ the equation
$$
 ax = x( \: {\rm Re}\,a  + |{\rm Im}\,a |i \: ) \eqno (3.6) 
$$
always has a nonzero solution$,$ namely$,$ $ a  \ \sim \  {\rm Re}\,a
+ |{\rm Im}\,a |i,$ and a  solution to Equation {\rm (3.6)} is
$$
  x = \lambda_1( \, |{\rm Im}\,a |i  + {\rm Im}\,a \, ) + \lambda_2
  [ \, |{\rm Im}\,a | -( {\rm Im}\,a )i \, ] \qquad   for \ all \ \
  \lambda_1, \  \lambda_2 \in {\cal R }.
$$
In particular$,$ if $ a = a_0 + a_1i $ with $ a_1 < 0$, then the general
solution to {\rm (3.6)} is
$$ 
x = x_2e_2 + \cdots +  x_7 e_7, \ \  for \ all \ x_2, \ \cdots, \
 x_7 \in {\cal R }.
$$}

The result in (3.6) can also be written as 
$a = x( \: {\rm Re}\,a  + |{\rm Im}\,a |i \: )x^{-1}$. Through it one can 
easily find out powers and $n$th roots of octonions. 

\medskip

\noindent {\bf  Theorem 3.3.}  \ {\em  Let $a = a_0 + a_1e_1 + \cdots +  a_7e_7 $
and $b = b_0 + b_1e_1 + \cdots +
 b_7e_7 $ be two given octonions. Then the equation
$$
 ax = \overline{x}b  \eqno (3.7) 
$$ 
has a nonzero solution for $ x \in {\cal O }$ if and only if  
$$
|a| = |b |. \eqno (3.8) 
$$ 
In that case$,$  if $ a + \overline{b} \neq 0,$ then {\rm (3.7)} has a
solution as follows
$$
 x = \lambda(  \overline{a} + b ),   \eqno (3.9)
$$ 
where $ \lambda \in {\cal R }$ is arbitrary$.$ In particular$,$
if $ a + \overline{b} = 0,$ i.e., $a_0  + b_0 =  0 $ and $ {\rm Im}\,a
= {\rm Im}\,b,$  then the general solution to {\rm (3.6)} is
$$
 x =  x_0 +  x_1e_1 + \cdots + x_7e_7, \eqno (3.10)
$$
where  $x_0$---$x_7 $ satisfy $  a_0x_0 - a_1x_1 - \cdots - a_7x_7 = 0.$  }

\medskip

\noindent {\bf Proof.} \  Suppose first that (3.7) has a nonzero solution
 $ x \in {\cal O }$. Then by (1.12) we get
$$ 
ax = xb \ \Longrightarrow \ |ax| = |xb| \   \Longrightarrow \
|a||x| = |x||b| \ \Longrightarrow \
 |a| = |b|. 
$$
Conversely, we let $ x_1 =\overline{a} + b$. Then
$$
ax_1- \overline{x_1}b  =  a(\overline{a} + b ) -
 ( a +  \overline{b})b =  a\overline{a} + ab - ab +
  \overline{b}b = |a|^2 - |b|^2.
$$   
Thus (3.9) is a solution to (3.7) under (3.8). If $ a + \overline{b}
 = 0$, then (3.7) is equivalent to
$ax +  \overline{ax} = 0$, i.e., $ {\rm Re\,}(ax)= 0$. Thus we have (3.10).
 \qquad  $\Box $ 

\medskip  

Based on the equation in (3.7) we can also define  the consimilarity
of octonions. Two octonions $ a $ and  $b $  are said to be {\em consimilar}
if there is a nonzero $  p \in {\cal O }$  such that
$ a = \overline{p}bp^{-1}$. By Theorem 3.3, we immediately know that
two octonions are  consimilar  if and only if their norms are identical.
Thus  the consimilarity defined here is also an equivalence relation on
octonions.

\medskip  

\noindent {\bf Corollary 3.4.} \ {\em  Any octonion $ a $ with $ a \notin
{\cal R }$ is  consimilar  to its norm,
namely, $ a =  \overline{p}|a|p^{-1}$, where $ p = |a| + \overline{a} $.
 } 

\medskip  

\noindent {\bf Corollary 3.5.} \ {\em Let  $a, \ b  \in {\cal O }$ be given  with $ a,
 \ b \notin {\cal R }.$ Then
 $a|a|^{-1} $ and $ b|b|^{-1} $ are consimilar. }

\medskip  

\noindent {\bf Corollary 3.6.}   \ {\em  Let $ a \in {\cal O }$ be given with $ a
\notin {\cal R }$.  Then the quadratic equation $ x^2 = a$
has two octonion solutions as follows
$$ 
x = \pm \frac{|a|^{\frac{1}{2}}( \, |a| + a \,)}{  | \, |a| + a \,|}
= \pm ( \, \lambda_0 + \lambda_1 a \,), \eqno (3.11)
$$
where $ \lambda_0 =   \frac{|a|^{\frac{3}{2}} }{  | \, |a| + a \,|}$ and $
 \lambda_1 = \frac{|a|^{\frac{1}{2}}}{  | \, |a| + a \,|}.$ } 

\medskip  

\noindent  {\bf Proof.} \  Follows from Corollary 3.4. \qquad  $\Box $ 

\medskip  

Based on the result in Corollary 3.6, solutions  can also be found for some other  quadratic  equations over ${\cal O}$ , such as,  $ xax = b, \ x^2 + bx + xb + c = 0 $ and  $x^2 + xb + c = 0 $ with $ c \in {\cal A}(b), $ the subalgebra generated by $ b $. 

\medskip

{\bf Theorem 3.7.} \ {\em  Let $ a, \ b  \in {\cal O }$ be given with $ a, \ b \notin {\cal R }$.  Then the equation $ \overline{x}ax = b$ is solvable if and only  if there is a $\lambda \in {\cal R }$ with $\lambda > 0$ such that 
$$ 
{\rm Re\,}a = \lambda {\rm Re\,}b, \ \ \ and \ \ \ |{\rm Im\,}a| 
= \lambda |{\rm Im \,}b|. \eqno (3.12)
$$ 
In that case$,$ a solution to $ \overline{x}ax = b$ can be written as 
$$ 
x = \frac{({\rm Im\,}a)p + p({\rm Im\,}b)}{ \sqrt{\lambda} \, |\, ({\rm Im\,}a)p + p({\rm Im\,}b) \,| }
,\eqno (3.13) 
$$ 
where $ p \in {\cal A}(a, \ b) $ is arbitrarily chosen such that 
$ ({\rm Im\,}a)p + p({\rm Im\,}b) \neq 0$.}  

\medskip

\noindent {\bf Proof.} \ Notice that $x^{-1} =  \overline{x}/|x|^2$ for a nonzero quaternion $x$.  Thus the equation $ \overline{x}ax = b$ can equivalently be written as 
$$ 
ax = \frac{x}{|x^2|}b. \eqno (3.14) 
$$ 
If (3.14) is solvable for $x$, then 
$$
{\rm Re\,}a = \frac{1}{|x|^2}{\rm Re\,}b, \ \ \ {\rm and} \ \ \ 
|{\rm Im\,}a| = \frac{1}{|x|^2}|{\rm Im\,}b|
$$
by Theorem 3.1(a), which implies (3.12). Conversely if (3.12) holds,  it is easy to verify that (3.13) satisfies (3.14). \qquad $\Box$ \\

\noindent {\bf 4. Similarity and consimilarity of sedenions}
 \\

\noindent The results in the above two sections can partly be extended
to the sedenion algebra ${\cal S }$.  

\medskip

\noindent {\bf Theorem 4.1.} \ {\em  Let $a = a_0 + a_1e_1 + \cdots +
 a_{15}e_{15}$ and $ b = b_0 + b_1e_1 + \cdots +  b_{15}e_{15}$
 be two given sedenions. If
$$
 a_0 = b_0, \qquad  and  \qquad | {\rm Im\,}a | = | {\rm Im\,}b |,
  \eqno (4.1)
 $$
and $  b \neq \overline{ a },$  then the linear equation
$$
 ax = xb   \eqno (4.2) 
$$ 
has a  solution as follows
$$
 x = \lambda ( \: {\rm Im\,}a  + {\rm Im\,}b\:), \eqno (4.3)
$$ 
where $ \lambda \in {\cal R }$ is arbitrary. If $ b = \overline{ a },$
 then the general solution to
Equation {\rm (4.2)} is
$$
 x =  x_1e_1 + x_2e_2  + \cdots +  x_{15}e_{15}, \eqno (4.4)
$$
where  $x_1$---$x_{15}$ satisfy $ a_1x_1 + a_2x_2  + \cdots +
 a_{15}x_{15} = 0.$  }

\medskip

\noindent {\bf Proof.}  \ Let $ x_1= {\rm Im\,}a  + {\rm Im\,}b$. Then it
 is easy to verify that
\begin{eqnarray*} 
ax_1- x_1b & = & ( a_0 + {\rm Im\,}a )( {\rm Im\,}a + {\rm Im}\,b )
 - ( {\rm Im}\,a + {\rm Im}\,b )( b_0 + {\rm Im}\,b) \\
 & = & ( a_0 -b_0 )x_1 + ({\rm Im\,}a )^2 + ( {\rm Im\,}a)({\rm Im\,}b ) - 
( {\rm Im\,}a)({\rm Im\,}b ) - ({\rm Im\,}b )^2 \\
& = &  ( a_0 -b_0 )x_1 + |{\rm Im}\,b|^2 - |{\rm Im}\,a|^2.    
\end{eqnarray*}   
Under (4.1), the right-hand side of the above equality is zero.
Thus (4.3) is a  nonzero solution of (4.2).
If $ b = \overline{ a },$ then (4.2) is $ ax = x\overline{a}$,
namely
$$ 
2 ({\rm Re\,}x) ( {\rm Im\,}a) +  ({\rm Im\,}a)({\rm Im\,}x) +
 ({\rm Im\,}x)({\rm Im\,}a) = 0,
$$ 
which is also equivalent to 
$$ 
{\rm Re\,}x = 0,  \ \ and  \ \   a_1x_1 + a_2x_2 + \cdots +
 a_{15}x_{15} = 0.
$$ 
Thus we have (4.4).  \qquad  $\Box $

\medskip

Since ${\cal S } $ is not a division algebra, it may occur
that there is a  nonzero $ x \in {\cal S }$ such that $ ax = 0$
and $xb= 0 $ for some $ a, \ b \in {\cal S }$. Thus (4.1) is not
a necessary condition for (4.2) to have a nonzero solution. Besides,
since $ {\cal S }$ is non-alternative, the sedenion
$ x_2 = |{\rm Im}a |\:|{\rm Im\,}b|  -( {\rm Im\,}a )({\rm Im\,}b) $
is no longer a solution to (4.2) under (4.1). Nevertheless, the similarity
concept can reasonably be established  for elements in ${\cal S }$,
according to the result in Theorem 4.1 as follows: two sedenions $ a $ and
$ b $ are said to be {\em similar} if ${\rm Re\,}a = {\rm Re\,}b$ and
$|{\rm Im \,}a | = |{\rm Im\,}b|$. Clearly the similarity defined above is also an equivalence relation for elements in {\cal S}. Moreover, we still have the
following. 

\medskip
  
\noindent {\bf Corollary 4.2.} \ {\em Let $ a \in { \cal S }$ given with
$ a \notin {\cal C}$. Then $a$ is similar to the complex number ${\rm Re}\,a  +
|{\rm Im}\,a |i $ and both of them satisfy the equality
$$
 ax = x( \, {\rm Re}\,a  + |{\rm Im}\,a |i \, ),   \eqno (4.5)
$$
where
$$
 x = \lambda(  \, |{\rm Im}\,a |i + {\rm Im}\,a \, ), \  \ \ \, \lambda \in {\cal R }.
$$
In particular$,$ if $ a = a_0 + a_1i $ with $ a_1 < 0,$ then the
general solution to {\rm (4.5)} is
$$ 
x = x_2e_2 + \cdots +  x_{15} e_{15}, \ \ \  for \ all \ \ \
 x_2, \ \cdots, \   x_{15} \in {\cal R }.
$$ }

\noindent {\bf Theorem 4.3.} \ {\em  Let $a = a_0 + a_1e_1 + \cdots +
  a_{15}e_{15} $ and $b = b_0 + b_1e_1 + \cdots + b_{15}e_{15} $
  be two given sedenions. If
$$
|a| = |b |, \eqno (4.6) 
$$ 
and $ a + \overline{b} \neq 0,$  then the equation
$$
 ax = \overline{x}b   \eqno (4.7) 
$$ 
has a solution as follows
$$
 x = \lambda(  \overline{a} + b ),  \eqno (4.8) 
$$ 
where $ \lambda \in {\cal R }$ is arbitrary. If $ a + \overline{b}
 = 0,$ then the general solution to {\rm (4.7)} is
$$
 x =  x_0 +  x_1e_1 + \cdots + x_{15}e_{15}, \eqno (4.9)
$$
where  $x_0$---$x_{15} $ satisfy $  a_0x_0 - a_1x_1 - \cdots -
 a_{15}x_{15} = 0.$  }  

\medskip

\noindent {\bf Proof.}  \ Let $ x_1 =\overline{a} + b$. Then
$$
ax_1- \overline{x_1}b  =  a(\overline{a} + b ) -
( a +  \overline{b})b =  a\overline{a} + ab - ab +
 \overline{b}b = |a|^2 - |b|^2.
$$   
Thus (4.8) is a solution to (4.7) under (4.6). If $ a +
\overline{b} = 0$, then (4.7) is equivalent to
$ax +  \overline{ax} = 0$, i.e., $ {\rm Re\,}(ax)= 0$.
Thus we have (4.9).     \qquad $\Box $ 

\medskip

Since ${\cal S } $ is not a division algebra, it may occur
that there is a nonzero $ x \in {\cal S }$ such that $ ax = 0$
and $\overline{x}b= 0 $ for certain $ a, \ b \in {\cal S }$.
Thus (4.6) is not a necessary condition for (4.7) to have a
nonzero solution. Nevertheless, we can reasonably introduce
consimilarity concept for elements in $ {\cal S }$ as follows:
two sedenions $a$ and $b$ are said to be {\em consimilar} if $ | a| = |b|$.
In that case, there is, by Theorem 4.3, an $ x \neq 0$ such that
$ ax = \overline{x}b$. Moreover, we still have the following
two results.

\medskip

\noindent {\bf Corollary 4.4.}  \ {\em  Any sedenion $ a \in {\cal S}$ with
$ a \notin {\cal R }$ is consimilar to its norm $|a|,$ and both of them
 satisfies the following equality
$$
 a =  \overline{p}|a|p^{-1},  \eqno (4.10)
$$
where $ p = |a| + \overline{a} $. } 

\medskip

\noindent {\bf Proof.} \  Follows from a direct verification. \qquad $\Box $ 

\medskip 

\noindent {\bf Theorem 4.5.} \ {\em  Let $ a \in {\cal S }$ given with $ a
\notin {\cal R }$.  Then the quadratic equation $x^2 = a $ has two
sedenion solutions as follows
$$ 
x = \pm \frac{|a|^{\frac{1}{2}}( \, |a| + a \,)}{  | \, |a| + a \,|}
= \pm ( \, \lambda_0 + \lambda_1 a \,),
$$
where $ \lambda_0 =   \frac{|a|^{\frac{3}{2}} }{  | \, |a| + a \,|}$
and $ \lambda_1 = \frac{|a|^{\frac{1}{2}}}{  | \, |a| + a \,|}.$ } 

\medskip

Correspondingly, solutions can also derived to the sedenion equations $ x^2 + bx + xb + c = 0 $ and $x^2 + xb + c = 0 $ with $ c \in {\cal A}(b),$ the subalgebra generated by $ b $.  But the solution of
  $ xax = b$ can not be derived from Theorem 4.5, because $xax = b$, in
  general, is not equivalent to $( ax )( ax ) = ab$ over ${\cal S }$.

\medskip

\noindent {\bf Remarks.} \ As is well known(see, e.g., \cite{AW}, \cite{Bro}, \cite{Mo}, \cite{Sc1}, \cite{Sc2}, \cite{We}), the  real Cayley-Dickson algebra $ {\cal A}_n$ when $ n \geq 4 $ is inductively defined by
adding a new generator
$\tau$ in ${\cal A}_{n-1}$. In that case, the elements in ${\cal A}_{n}$ take form 
$$
a = a' + a''\tau,
$$
where $a', \  a'' \in  {\cal A}_{n-1}$.  The {\em conjugate} $ a$ is 
defined to be 
$$ 
\overline{a} = \overline{a'} - a'' \tau = {\rm Re\,}a -  {\rm Im\,}a.
$$
The {\em norm} of $ a $ is defined to be 
$$  
|a | = \sqrt{a\overline{a}} = \sqrt{\overline{a}a} = \sqrt{ |a'|^2 + |a''|^2}. 
$$
The addition and multiplication for
any  $ a = a' + a''\tau, \ b = b' + b''\tau \in {\cal A}_{n}$ are 
$$ 
a  + b = ( \, a' + a''\tau \,)  + ( \, b' + b''\tau \, ) =  (\, a' + b'\,) +
( \, a'' +  b'' \,)\tau,
$$  
$$ 
ab = ( \,a' + a''\tau \, )( \, b' + b''\tau \,) = ( \, a'b' - \overline{b''}a'' \, ) +
( \, b''a' +  a''\overline{b'} \,)\tau.  
$$ 
In that case, ${\cal A}_{n}$ is spanned as a  $2^n$-dimensional 
non-commutative, non-alternative, non-composition and non-division algebra
over ${\cal R}$, but it is still a power-associative, flexible and quadratic
 algebra over $ {\cal R}$ when $ n \geq 4$. The algebraic properties of
 $ {\cal A}_n$ with $ n \geq 4 $  are much similar to those of
 $ {\cal A}_4 = {\cal S}$. Therefore the results in Section 4 can directly be
 extended to elements in $ {\cal A}_n$. For simplicity, we do not intend to list 
them here.

\bigskip
\end{document}